\documentclass[useAMS,usenatbib,usegraphicx]{mn2e}

\usepackage{color}
\usepackage{natbib, aas_macros}
\title[LoCuSS Hydrostatic bias]{LoCuSS: Testing hydrostatic equilibrium in galaxy clusters}
\author[ G.\ P.\ Smith et al.]
       {G.\ P.\ Smith,$\!^1$\thanks{E-mail: gps@star.sr.bham.ac.uk}
         P.\ Mazzotta,$\!^2$
         N.\ Okabe,$\!^{3,4,5}$
         F.\ Ziparo,$\!^1$
         S.\ L.\ Mulroy,$\!^1$
         A.\ Babul,$\!^6$\newauthor
         A.\ Finoguenov,$\!^7$
         I.\ G.\ McCarthy,$\!^8$
         M.\ Lieu,$\!^1$
         Y.\ Bah\'e,$\!^9$
         H.\ Bourdin,$\!^2$
         A.\ E.\ Evrard,$\!^{10,11}$\newauthor
         T.\ Futamase,$\!^{12}$
         C.\ P.\ Haines,$\!^{1,13,14}$
         M.\ Jauzac,$\!^{15,16}$
         D.\ P.\ Marrone,$\!^{13}$
         R.\ Martino,$\!^2$\newauthor
         P.\ E.\ May,$\!^1$
         J.\ E.\ Taylor,$\!^{17}$
         K.\ Umetsu $\!^{18}$\\\\
         $^1$ School of Physics and Astronomy, University of Birmingham, Birmingham, B15 2TT, England.\\
         $^2$ Dipartimento di Fisica, Universit\`a degli Studi di Roma `Tor Vergata', via della Ricerca Scientifica 1, 00133 Roma, Italy.\\
         $^3$ Department of Physical Science, Hiroshima University, 1-3-1
         Kagamiyama, Higashi-Hiroshima, Hiroshima 739-8526, Japan.\\
         $^4$ Hiroshima Astrophysical Science Center, Hiroshima University, Higashi-Hiroshima, Kagamiyama 1-3-1, 739-8526, Japan.\\
         $^5$ Kavli Institute for the Physics and Mathematics of the Universe (WPI), Todai Institutes for Advanced Study, University of Tokyo, 5-1-5 Kashiwanoha, Kashiwa, Chiba 277-8583, Japan.\\
         $^6$ Department of Physics and Astronomy, University of Victoria, 3800 Finnerty Road, Victoria, BC V8P 1A1, Canada\\
         $^7$ Department of Physics, University of Helsinki, Gustaf H\"allstr\"omin katu 2a, 00014, Helsinki, Finland\\
         $^8$ Astrophysics Research Institute, Liverpool John Moores University, 146 Brownlow Hill, Liverpool L3 5RF, England\\
         $^9$ Max-Planck-Institut f\"ur Astrophysik, Karl-Schwarzschild Str. 1, D-85748 Garching, Germany\\
         $^{10}$ Department of Physics and Michigan Center for Theoretical Physics, University of Michigan, Ann Arbor, MI 48109, USA\\
         $^{11}$ Department of Astronomy, University of Michigan, Ann Arbor, MI 48109, USA\\
         $^{12}$ Astronomical Institute, Tohoku University, Aramaki, Aoba, Sendai 980-8578, Japan\\
         $^{13}$ Steward Observatory, University of Arizona, 933 North Cherry Avenue, Tucson, AZ 85721, USA\\
         $^{14}$ Departamento de Astronom\'ia, Universidad de Chile, Casilla 36-D, Correo Central, Santiago, Chile\\
         $^{15}$ Institute for Computational Cosmology, Durham University, South Road, Durham DH1 3LE, England\\
         $^{16}$ Astrophysics and Cosmology Research Unit, School of Mathematical Sciences, University of KwaZulu-Natal, Durban 4041, South Africa\\
         $^{17}$ Department of Physics and Astronomy, University of Waterloo, 200 University Avenue West, Waterloo, ON N2L 3G1, Canada\\
         $^{18}$ Academia Sinica Institute of Astronomy and Astrophysics (ASIAA), P.\ O.\ Box 23-141, Taipei 10617, Taiwan
}

\begin{document}

\date{\today}

\pagerange{\pageref{firstpage}--\pageref{lastpage}} \pubyear{2015}

\maketitle

\label{firstpage}

\newcommand{\simgt}{\lower.5ex\hbox{$\; \buildrel > \over \sim \;$}}
\newcommand{\simlt}{\lower.5ex\hbox{$\; \buildrel < \over \sim \;$}}
\newcommand{\tp}{\hspace{-1mm}+\hspace{-1mm}}
\newcommand{\tm}{\hspace{-1mm}-\hspace{-1mm}}
\newcommand{\gIII}{I\hspace{-.3mm}I\hspace{-.3mm}I}
\newcommand{\bmf}[1]{\mbox{\boldmath$#1$}}
\def\bbeta{\mbox{\boldmath $\beta$}}
\def\btheta{\mbox{\boldmath $\theta$}}
\def\bnabla{\mbox{\boldmath $\nabla$}}
\def\bk{\mbox{\boldmath $k$}}
\newcommand{\cA}{{\cal A}}
\newcommand{\cD}{{\cal D}}
\newcommand{\cF}{{\cal F}}
\newcommand{\cG}{{\cal G}}
\newcommand{\trQ}{{\rm tr}Q}
\newcommand{\Real}[1]{{\rm Re}\left[ #1 \right]}
\newcommand{\paren}[1]{\left( #1 \right)}
\newcommand{\red}{\textcolor{red}}
\newcommand{\blue}{\textcolor{blue}}
\newcommand{\norv}[1]{{\textcolor{blue}{#1}}}
\newcommand{\norvnew}[1]{{\textcolor{red}{#1}}}

\def\hkpc{\mathrel{h^{-1}{\rm kpc}}}
\def\hMpc{\mathrel{h^{-1}{\rm Mpc}}}
\def\Mpc{\mathrel{\rm Mpc}}
\def\Mvir{\mathrel{M_{\rm vir}}}
\def\cvir{\mathrel{c_{\rm vir}}}
\def\rvir{\mathrel{r_{\rm vir}}}
\def\Dvir{\mathrel{\Delta_{\rm vir}}}
\def\rsc{\mathrel{r_{\rm sc}}}
\def\rhoc{\mathrel{\rho_{\rm crit}}}
\def\Msol{\mathrel{M_\odot}}
\def\hMsol{\mathrel{h^{-1}M_\odot}}
\def\h70Msol{\mathrel{h_{70}^{-1}M_\odot}}
\def\ergs{\mathrel{\rm erg\,s^{-1}}}
\def\Mgas{\mathrel{M_{\rm gas}}}
\def\Mhse{\mathrel{M_{\rm HSE}}}
\def\Mhsei{\mathrel{M_{\rm HSE,i}}}
\def\Mx{\mathrel{M_{\rm X}}}
\def\Mxi{\mathrel{M_{\rm X,i}}}
\def\Mp{\mathrel{M_{\rm Planck}}}
\def\Mpi{\mathrel{M_{\rm Planck,i}}}
\def\Mwl{\mathrel{M_{\rm WL}}}
\def\Mwli{\mathrel{M_{\rm WL,i}}}
\def\Mfh{\mathrel{M_{500}}}
\def\rfh{\mathrel{r_{500}}}
\def\rwlfh{\mathrel{r_{\rm WL,500}}}
\def\Tx{\mathrel{T_X}}
\def\Om{\mathrel{\Omega_{\rm M}}}
\def\Ol{\mathrel{\Omega_\Lambda}}
\def\keV{\mathrel{\rm keV}}
\def\kpc{\mathrel{\rm kpc}}
\def\kms{\mathrel{\rm km\,s^{-1}}}
\def\ls{\mathrel{\hbox{\rlap{\hbox{\lower4pt\hbox{$\sim$}}}\hbox{$<$}}}}
\def\gs{\mathrel{\hbox{\rlap{\hbox{\lower4pt\hbox{$\sim$}}}\hbox{$>$}}}}
\def\ds{\mathrel{D_{\rm S}}}
\def\dls{\mathrel{D_{\rm LS}}}
\def\dsi{\mathrel{D_{{\rm S},i}}}
\def\dlsi{\mathrel{D_{{\rm LS},i}}}
\def\dsj{\mathrel{D_{{\rm S},j}}}
\def\dlsj{\mathrel{D_{{\rm LS},j}}}
\def\dsi{\mathrel{D_{{\rm S},i}}}
\def\zs{\mathrel{z_{S}}}
\def\ks{\mathrel{\rm ksec}}
\def\betaP{\mathrel{\beta_{\rm P}}}
\def\betaX{\mathrel{\beta_{\rm X}}}

\addtolength\topmargin{-12mm}

\begin{abstract}
We test the assumption of hydrostatic equilibrium in an X-ray
luminosity selected sample of 50 galaxy clusters at $0.15<z<0.3$ from
the Local Cluster Substructure Survey (LoCuSS).  Our weak-lensing
measurements of $M_{500}$ control systematic biases to sub-4 per cent,
and our hydrostatic measurements of the same achieve excellent
agreement between \emph{XMM-Newton} and \emph{Chandra}.  The mean
ratio of X-ray to lensing mass for these 50 clusters is
$\betaX=0.95\pm0.05$, and for the 44 clusters also detected by
\emph{Planck}, the mean ratio of \emph{Planck} mass estimate to LoCuSS
lensing mass is $\betaP=0.95\pm0.04$.  Based on a careful
like-for-like analysis, we find that LoCuSS, the Canadian Cluster
Comparison Project (CCCP), and Weighing the Giants (WtG) agree on
$\betaP\simeq0.9-0.95$ at $0.15<z<0.3$.  This small level of
hydrostatic bias disagrees at $\sim5\sigma$ with the level required to
reconcile \emph{Planck} cosmology results from the cosmic microwave
background and galaxy cluster counts.
\end{abstract}


\makeatletter
\def\doi{\begingroup
  \let\do\@makeother \do\\\do\$\do\&\do\#\do\^\do\_\do\%\do\~
  \@ifnextchar[%]
    {\@doi}
    {\@doi[]}}
\def\@doi[#1]#2{%
  \def\@tempa{#1}%
  \ifx\@tempa\@empty
    \href{http://dx.doi.org/#2}{doi:#2}%
  \else
    \href{http://dx.doi.org/#2}{#1}%
  \fi
  \endgroup
}

\def\eprint#1#2{%
  \@eprint#1:#2::\@nil}
\def\@eprint@arXiv#1{\href{http://arxiv.org/abs/#1}{{\tt arXiv:#1}}}
\def\@eprint@dblp#1{\href{http://dblp.uni-trier.de/rec/bibtex/#1.xml}{dblp:#1}}
\def\@eprint#1:#2:#3:#4\@nil{%
  \def\@tempa{#1}%
  \def\@tempb{#2}%
  \def\@tempc{#3}%
  \ifx\@tempc\@empty
    \let\@tempc\@tempb
    \let\@tempb\@tempa
  \fi
  \ifx\@tempb\@empty
    \def\@tempb{arXiv}%
  \fi
  \@ifundefined{@eprint@\@tempb}
    {\@tempb:\@tempc}
    {\expandafter\expandafter\csname @eprint@\@tempb\endcsname\expandafter{\@tempc}}
}

\def\mniiiauthor#1#2#3{%
  \@ifundefined{mniiiauth@#1}
    {\global\expandafter\let\csname mniiiauth@#1\endcsname\null #2}
    {#3}}

\makeatother

\section{Introduction}\label{sec:intro}

Accurate measurement of systematic biases in galaxy cluster masses is
fundamental to cosmological exploitation of galaxy clusters, as has
been highlighted recently by \citet{Planck15SZcos}.  Much attention
has focused on the systematic biases in the respective mass
measurement techniques, principally via weak-lensing
\citep[e.g.][]{Okabe13, Applegate14, Hoekstra15, Okabe15} and X-ray
\citep[e.g.][]{Rasia06, Nagai07, Meneghetti10a, Rasia12, Martino14}
methods.  Specifically, comparing lensing- and X-ray-based mass
measurements tests the hydrostatic equilibrium assumption that
underpins the X-ray-based mass measurements \citep[e.g.][]{Miralda95,
  Allen98, Smith01a, Smith05a, Mahdavi08, Zhang10, Richard10,
  Mahdavi13, Israel14}.

Our goal is to assess the implications of the new LoCuSS weak-lensing
mass calibration \citep{Okabe15,Ziparo15} for hydrostatic bias and
thus systematic uncertainties in cluster cosmology results.  We
combine \citeauthor{Okabe15}'s masses with hydrostatic masses from
\citet{Martino14}.  Both \citeauthor{Okabe15} and
\citeauthor{Martino14} control systematic biases in their respective
mass measurements at sub-$4$ per cent.  They are arguably the most
accurate cluster mass measurements available to date.  We also use
mass estimates from \citet{Planck15SZsample} that assume hydrostatic
equilibrium, via an X-ray scaling relation and measurements of the
integrated Compton $Y$ parameter from \emph{Planck} survey data.  We
describe our analysis and results in Section~\ref{sec:results},
discuss our results in Section~\ref{sec:discuss}, and conclude in
Section~\ref{sec:conc}.  We assume $H_0=70\kms\Mpc^{-1}$, $\Om=0.3$
and $\Ol=0.7$ throughout.

\section{Analysis and results}\label{sec:results}

\subsection{Sample and mass measurements}\label{sec:mass}

The sample comprises 50 clusters from the \emph{ROSAT} All-sky Survey
catalogues \citep{Ebeling98,Ebeling00,Boehringer04} that satisfy:
$-25^\circ<\delta<+65^\circ$, $n_H\le7\times10^{20}{\rm cm}^{−2}$,
$0.15\le z\le0.3$, $L_X[0.1-2.4\,{\rm
    keV}]/E(z)\ge4.1\times10^{44}\ergs$, where
$E(z)=\sqrt{\Om(1+z)^3+\Ol}$. The clusters are therefore selected
purely on $L_X$, ignoring other physical parameters.  We focus on
measurements of $\Mfh$, defined as the mass enclosed within $\rfh$,
i.e.\ the radius within which the mean density of the cluster is 500
times the critical density of the universe ($\rhoc$).  $\Mfh$ for a
cluster at a redshift of $z$ is therefore:
${\Mfh}=500{\rhoc}(z)\,4{\pi}{\rfh}^3/3$.

We use weak-lensing masses from \cite[][see also Ziparo et
  al.\ 2015]{Okabe15}.  The two largest systematic biases in these
weak-lensing masses are shear calibration ($3$ per cent) and
contamination of background galaxy catalogues ($1$ per cent).  The
former calibration is derived from extensive image simulations,
including shears up to $g\simeq0.3$; the latter is based on selecting
galaxies redder than the red sequence of cluster members using a
radially-dependent colour-cut.  \citeauthor{Okabe15} also used full
cosmological hydrodynamical numerical simulations
\citep{McCarthy14,LeBrun14} to calibrate systematic biases in mass
modeling to sub-1 per cent.  In this article we use weak-lensing mass
measurements calculated after correcting for the shape measurements
and contamination biases -- see \citeauthor{Okabe15}'s Table~A.1.

We use hydrostatic masses from \citet{Martino14}, who modelled X-ray
observations of the clusters assuming that the X-ray emitting cluster
gas is in hydrostatic equilibrium with the cluster potential.  Forty
three had been observed by \emph{Chandra} and 39 with
\emph{XMM-Newton}.  For the 21 clusters observed by both, the average
ratio of \emph{Chandra} to \emph{XMM-Newton} hydrostatic mass was
$1.02\pm0.05$ with an intrinsic scatter of $\sim8$ per cent.  We use
hydrostatic $\Mfh$ from Table~2 of \citeauthor{Martino14}, adopting
masses from \emph{Chandra} where available, and otherwise from
\emph{XMM-Newton} data.  We add 8 per cent systematic uncertainty in
quadrature to the statistical error on hydrostatic mass to account for
the intrinsic scatter noted above.  Note that \citeauthor{Martino14}
use data from ACIS-I and ACIS-S on \emph{Chandra} and EPIC (including
both PN and MOS) on \emph{XMM-Newton}.

We obtain estimates of $\Mfh$ from \cite{Planck15SZsample} for 44
clusters.  These masses are based on measurements of the spherical
Compton $Y$ measurement from the millimetre wave data, and a
relationship between $Y_X$ and $\Mfh$ derived from X-ray observations
of a sample of 20 clusters at $z<0.2$ selected to have ``relaxed''
X-ray morphology, where $Y_X$ is the iteratively defined
pseudo-pressure of the X-ray emitting gas, ${Y_X}\equiv{\Mgas}.{\Tx}$
\cite[][]{Arnaud07,Arnaud10}.  As such, the \emph{Planck} mass
estimates assume the clusters are in hydrostatic equilibrium.

\subsection{Method of calculation}\label{sec:method}

\begin{figure*}
\includegraphics[width=0.9\hsize]{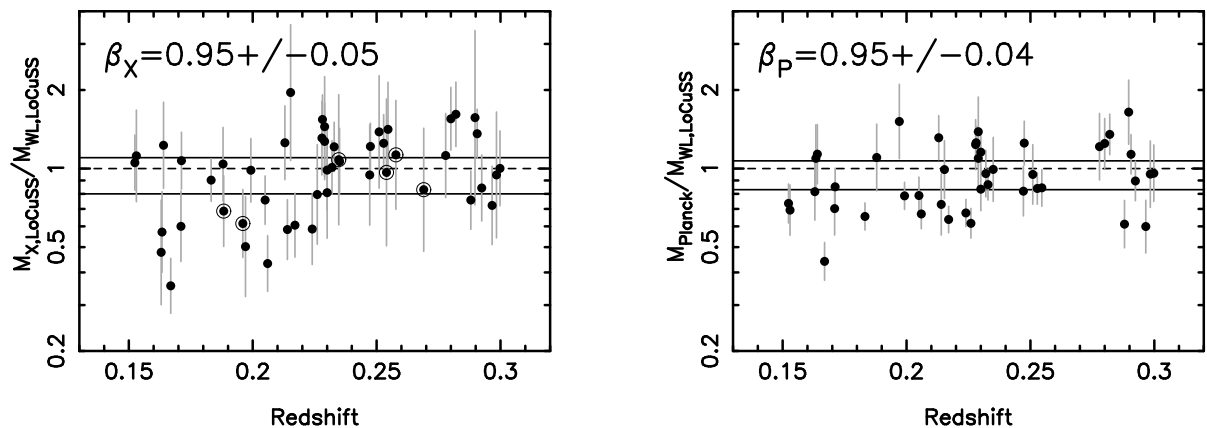}
\caption{{\sc Left} -- Ratio of X-ray-based hydrostatic mass to
  weak-lensing mass versus redshift for the 50 clusters in the LoCuSS
  sample, adding an open circle around clusters not detected by
  \emph{Planck}.  {\sc Right} -- Ratio of \emph{Planck} mass estimate
  to weak-lensing mass versus redshift for 44 clusters in the LoCuSS
  sample.  {\sc Both} -- The Pearson correlation coefficient for mass
  ratio versus redshift confirms that the \emph{possible} trend of
  mass ratio with redshift seen by eye in these panels is
  statistically insignificant.  The horizontal dashed lines mark
  $\beta=1$, and the solid lines show the $\pm3\sigma$ confidence
  interval on $\betaX$ (left; \S\ref{sec:mwlmhse}) and $\betaP$
  (right; \S\ref{sec:locussplanck}).}
\label{fig:betaz}
\end{figure*}

We define $\beta$ as the geometric mean ratio of the hydrostatic mass,
$\Mhse$, to the weak-lensing mass, $\Mwl$, for a sample of $n$
clusters:
\begin{equation}
\beta=\exp{\left[\frac{\sum_{i=1}^{n}w_i\,\ln\!\left(\frac{\Mhsei}{\Mwli}\right)}{\sum_{i=1}^{n}w_i}\right]},
\label{eqn:beta}
\end{equation}
where $w_i$ is the weight attached to each cluster.  We calculate the
uncertainty on $\beta$ as the standard deviation of the geometric
means of 1000 bootstrap samples each numbering $n$ clusters.
Measurements of $\beta$ based on direct measurement of $\Mhse$ from
X-ray data are denoted as $\betaX$, and measurements based on
\emph{Planck} mass estimates are denoted as $\betaP$.  

We aim to maximize sensitivity of the weights, $w_i$, to data quality,
and minimize sensitivity to physical properties and/or geometry of the
clusters.  When calculating $\betaX$ we adopt the reciprocal of the
sum of the squares of the fractional error on X-ray-based $\Mhse$
(denoted here explicitly as $\Mx$) and the absolute error on $\Mwl$:
\begin{equation}
w_i=\left[\left(\frac{\delta{\Mxi}/{\Mxi}}{\langle\delta{\Mx}/{\Mx}\rangle}\right)^2+\left(\frac{\delta{\Mwli}}{\langle\delta{\Mwl}\rangle}\right)^2\right]^{-1}
\label{wiX}
\end{equation}
The weighting with respect to the hydrostatic masses reflects the fact
the absolute error on $\Mx$ is tightly correlated with $\Mx$
itself.  This is because the X-ray spectra of more massive (hotter)
clusters contain less emission features than spectra of cooler
clusters, thus making hydrostatic mass measurements intrinsically less
precise for hotter clusters despite them being brighter.  In contrast
the fractional error on $\Mx$ is not a strong function of $\Mx$,
and so the mass dependence of the weighting scheme is significantly
reduced.  The weighting with respect to the weak-lensing masses
reflects the fact that the absolute error on $\Mwl$ traces the
weak-lensing data quality more faithfully than the fractional error on
$\Mwl$.  Indeed, given the uniformity of our weak-lensing data
\citep{Okabe15}, the fractional error would up-weight clusters with
large values of $\Mwl$, thus biasing our results to clusters with
large masses and/or that are observed at small angles with respect to
their major axis \citep{Meneghetti10a}.  The latter effect would
introduce a geometric bias into our results.  When calculating
$\betaP$ we adopt the reciprocal of the sum of the squares of the
absolute errors on $\Mp$ and $\Mwl$:
\begin{equation}
w_i=\left[\left(\frac{\delta{\Mpi}}{\langle\delta{\Mp}\rangle}\right)^2+\left(\frac{\delta{\Mwli}}{\langle\delta{\Mwl}\rangle}\right)^2\right]^{-1}
\label{wiP}
\end{equation}
The weighting with respect to the Planck mass estimates follows a
similar motivation to that described above for the weak-lensing
masses.  

\subsection{Comparing LoCuSS weak-lensing and X-ray masses}\label{sec:mwlmhse}

We compare weak-lensing masses with X-ray masses, with each computed
within their independently derived $\rfh$ (Figure~\ref{fig:betaz},
left panel), obtaining $\betaX=0.95\pm0.05$.  Arguably a more accurate
calculation uses hydrostatic and weak-lensing masses measured within
the same radius.  We therefore recalculate $\betaX$ based on X-ray and
lensing masses both computed within the weak-lensing-based $\rfh$
(hereafter $\rwlfh$), obtaining $\betaX=0.87\pm0.04$, $1.2\sigma$
lower than the former measurement, however note that adopting $\rwlfh$
as the radius for both masses introduces a covariance that we have
neglected in our calculation.

\subsection{Comparing LoCuSS weak-lensing masses and Planck mass estimates}\label{sec:locussplanck}

We compare weak-lensing mass measurements with the \emph{Planck} mass
estimates to compute $\betaP$ (Figure~\ref{fig:betaz}, right panel),
obtaining $\betaP=0.95\pm0.04$, in excellent agreement with $\betaX$
(\S\ref{sec:mwlmhse}).  Note that the apertures within which our
weak-lensing masses are computed are independent of the apertures used
by \citet{Planck15SZsample} when calculating the \emph{Planck} mass
estimates.  We double check the consistency between $\betaX$ and
$\betaP$ by repeating the X-ray/lensing comparison
(\S\ref{sec:mwlmhse}) for the 44 clusters detected by \emph{Planck}
and considered in this section, obtaining $\betaX=0.97\pm0.06$.  The
agreement between $\betaX$ and $\betaP$ is therefore not sensitive to
the six clusters that have not been detected by \emph{Planck}.

\section{Discussion}\label{sec:discuss}

We now compare our results with previous observational studies, noting
in passing that our measurements of hydrostatic bias are in line with
numerous cosmological numerical hydrodynamical simulations
\citep[e.g.][]{Nagai07, Meneghetti10a, Rasia12, LeBrun14}.

\subsection{Comparison with pointed X-ray surveys}\label{sec:compareX}

\citet{Martino14} compared their hydrostatic masses (used in this
letter) with LoCuSS weak-lensing masses \citep{Okabe10a,Okabe13},
obtaining $\betaX\simeq0.93$.  This result is fully consistent with
our $\betaX=0.95\pm0.05$, that uses the new LoCuSS weak-lensing masses
from \citet{Okabe15}.

The Canadian Cluster Comparison Project (CCCP) obtained
$\betaX=0.88\pm0.05$ with both hydrostatic and weak-lensing masses
measured within $\rwlfh$ \citep{Mahdavi13}.  \citet{Hoekstra15}
updated the CCCP weak-lensing masses, reporting masses
($\Mwl(<r_{500})$) on average $19$ per cent higher than
\citet{Hoekstra12} and \citet{Mahdavi13}.  Applying a factor 1.19
``correction'' to the denominator of the CCCP $\betaX$ implies
$\betaX\simeq0.74$.  However we note that \citet{Martino14} found that
Mahdavi et al.'s hydrostatic masses are on average $\sim14$ per cent
lower than LoCuSS hydrostatic masses for 21 clusters in common (see
Martino et al.\ for details).  Applying a further factor 1.14
correction to the numerator brings CCCP up to $\betaX\simeq0.84$, in
agreement with our $\betaX=0.87\pm0.04$ (\S\ref{sec:mwlmhse}).

\citet{Israel14} considered eight clusters at $z\simeq0.5$ from the
400d survey, obtaining $\betaX=0.92^{+0.09}_{-0.08}$, in good
agreement with our measurements.  Note that this is based on the first
line of their Table~2, which gives the most like-for-like comparison
with our methods.

After we submitted this letter \citet{Applegate15} posted a preprint
that compares weak-lensing and hydrostatic mass measurements within
X-ray-based $r_{2500}$ for a sample of 12 ``relaxed'' clusters.
Detailed comparison of their results with ours is hindered by the
absence of individual cluster masses in Applegate et al., and their
small sample.  Their main result is a ratio of weak-lensing mass to
hydrostatic mass within $r_{2500}$ of $0.96\pm0.13$.  They also
comment that they obtain a ratio of $1.06\pm0.13$ at $r_{500}$.  We
repeat our calculation of $\betaX$ described at the end of
\S\ref{sec:mwlmhse} within matched apertures with weak-lensing mass as
the numerator and hydrostatic mass as the denominator, obtaining a
weak-lensing to hydrostatic mass of $1.15\pm0.04$ at $r_{500}$.

\subsection{Comparison with Sunyaev-Zeldovich effect surveys}\label{sec:compareP}

Weighing the Giants (WtG) and CCCP have reported $\betaP=0.70\pm0.06$
and $\betaP=0.76\pm0.08$ respectively
\citep{vonderLinden14,Hoekstra15}, both based on the
\citet{Planck13SZsample} masses.  These measurements are lower than
our $\betaP=0.95\pm0.04$ at $3.5\sigma$ and $2.1\sigma$ respectively.

We apply our methods, including absolute mass errors weighting
(\S\ref{sec:method}), to the clusters and masses used by
\citet{vonderLinden14}, obtaining $\betaP=0.80\pm0.07$.
\citeauthor{vonderLinden14} do not state explicitly their method of
calculation, however if we weight uniformly then we obtain
$\betaP=0.69\pm0.07$, in agreement with them.  Next, we update the WtG
results to the \citet{Planck15SZsample} measurements of $\Mp$,
obtaining slightly higher values: $\betaP=0.72\pm0.07$ and
$\betaP=0.83\pm0.07$ for uniform and absolute mass error weighting
respectively.  Splitting the clusters into two redshift bins, with the
lower redshift bin matching LoCuSS, and again using absolute mass
error weighting, we obtain $\betaP(z<0.3)=0.90\pm0.09$ and
$\betaP(z>0.3)=0.71\pm0.07$.  This is consistent with our results at
$z<0.3$, and suggests $\betaP$ might be a function of redshift.

We also apply our methods to the clusters and masses considered by
\citet{Hoekstra15}, obtaining $\betaP=0.83\pm0.07$.  We reproduce the
published CCCP result if we weight the clusters uniformly, in which
case we obtain $\betaP=0.77\pm0.07$.  Updating to the
\citet{Planck15SZsample} masses, gives a slightly higher value of
$\betaP=0.85\pm0.08$ (using absolute mass error weights).  So far we
have followed \citeauthor{Hoekstra15} in using their deprojected
aperture mass measurements.  However, both LoCuSS and WtG obtain
masses by fitting an NFW model to the shear profile.  To obtain a
like-for-like comparison we therefore use \citeauthor{Hoekstra15}'s
NFW-based masses, the \cite{Planck15SZsample} masses, and absolute
mass error weights, obtaining $\betaP=0.92\pm0.08$.  Finally, we split
the CCCP sample into two redshift bins, as above, and find
$\betaP(z<0.3)=0.96\pm0.09$ and $\betaP(z>0.3)=0.61\pm0.09$.  This is
consistent with our results at $z<0.3$, again suggesting $\betaP$
depends on redshift.

After we submitted this letter \citet{Battaglia15b} reported
weak-lensing follow up of the Atacama Cosmology Telescope (ACT)
thermal Sunyaev-Zeldovich (SZ) cluster sample.  They commented that
WtG and CCCP measurements of $\betaP\simeq0.7-0.8$ may be biased
\emph{high} because clusters that are not detected by \emph{Planck}
are excluded from their calculations.  They estimated the possible
bias by assigning to the non-detections a mass equal to the
\emph{Planck} $5\sigma$ detection threshold and thus including these
clusters in the calculations of $\betaP$.  They found that this
reduces the CCCP and WtG $\betaP$ values by $\sim0.06$ and $\sim0.16$
respectively.  We expect any bias of this nature to be small in our
analysis because only six clusters from our sample of fifty are not
detected by \emph{Planck}.  Nevertheless, we perform the calculations
outlined by \citeauthor{Battaglia15b} and successfully reproduce their
values for WtG and CCCP.  We then estimated the possible bias in our
results, and find that including the 6 non-detections reduces our
measurement of $\betaP$ by just $\sim0.04$.  We also estimate the bias
for WtG and CCCP using just their clusters at $z<0.3$, and obtain
$\sim0.04$.  Biases caused by excluding \emph{Planck} non-detections
appear to dominate neither our results nor comparison with WtG and
CCCP at $z<0.3$.

\section{Conclusions and perspective on ``Planck Cosmology''}\label{sec:conc}

We have used three sets of independent mass measurements to develop a
consistent picture of the departures from hydrostatic equilibrium in
the Local Cluster Substructure Survey (LoCuSS) sample of 50 clusters
at $0.15\le z\le0.3$.  These clusters were selected purely on their
X-ray luminosity, declination, and line of sight hydrogen column
density.  The mass measurements comprise weak-lensing masses
\citep{Okabe15,Ziparo15}, direct measurements of hydrostatic masses
using X-ray observations \citep{Martino14}, and estimated hydrostatic
masses from \citet{Planck15SZsample}.  The main strength of our
results is the careful analysis of systematic biases in the
weak-lensing and hydrostatic mass measurements referred to above, and
summarized in \S\ref{sec:mass}.

We obtain excellent agreement between our X-ray-based and
\emph{Planck}-based tests of hydrostatic equilibrium, with
$\betaX=0.95\pm0.05$ (\S\ref{sec:mwlmhse}) and $\betaP=0.95\pm0.04$
(\S\ref{sec:locussplanck}).  The masses used for these calculations
are measured within independently derived estimates of $\rfh$.  We
also remeasured $\betaX$ using X-ray masses measured within the
weak-lensing-based $\rfh$, obtaining $\betaX=0.87\pm0.04$
(\S\ref{sec:mwlmhse}), suggesting that the actual level of hydrostatic
bias, of astrophysical interest, might be slightly larger than
inferred from the calculations based on independent measurement
apertures.

Our measurement of $\betaP$ is larger (implying smaller hydrostatic
bias) than recent results from the WtG and CCCP surveys
\citep{vonderLinden14,Hoekstra15} at $3.5\sigma$ and $2.1\sigma$
respectively (\S\ref{sec:compareP}).  However if we restrict the WtG
and CCCP sample to the same redshift range as LoCuSS ($0.15<z<0.3$),
use a consistent method to calculate $\betaP$ (\S\ref{sec:method}),
and incorporate up to date \emph{Planck} mass estimates
\citep{Planck15SZsample} into the WtG and CCCP calculations, we obtain
$\betaP(z<0.3)=0.90\pm0.09$ and $\betaP(z<0.3)=0.96\pm0.09$
respectively.  This highlights that the previously reported low values
of $\betaP$ appear to be dominated by clusters at $z>0.3$, with
$\betaP(z>0.3)\sim0.6-0.7$.  We also note that estimates of bias in
$\betaP$ caused by excluding clusters not detected by \emph{Planck}
\citep{Battaglia15b} are $\sim0.04$ for clusters at $z<0.3$, and
$\gs0.1$ at $z>0.3$, in the sense that these biases reduce $\betaP$.
In short, any bias appears to be sub-dominant to statistical
uncertainties at $z<0.3$, that is the main focus of this letter.

We are therefore lead to a view that $\betaP\simeq0.9-0.95$ at $z<0.3$
and $\betaP\ls0.6$ at $z>0.3$.  The very low value at $z>0.3$ could be
caused by systematic biases in mass measurements that relate to
observational or measurement effects, and not to the validity of
hydrostatic equilibrium.  It is plausibe that systematic biases in
weak-lensing mass measurements are better controlled at $z<0.3$ than
at $z>0.3$, because for observations to fixed photometric depth, the
sensitivity of the weak-lensing mass measurements to the accuracy of
the redshift distribution of the background galaxies increases with
cluster redshift.  It would also be interesting to consider the
possibility of redshift-dependent biases in the \emph{Planck} mass
estimates.

Our results imply a hydrostatic bias parameter, $(1-b)$, at the upper
end of the range of values considered as a prior by
\citet{Planck15SZcos} for their cluster cosmology analysis.
Intriguingly, our measurements are compatible with the CMB lensing
constraints of $(1-b)=1.01^{+0.24}_{-0.16}$ \citep{Melin15}, although
the uncertainties on this pioneering measurement were admittedly
large.  On the other hand, our measurements disagree at $\sim5\sigma$
with the value of $(1-b)=0.58\pm0.04$ computed by
\citet{Planck15SZcos} as being required to reconcile the \emph{Planck}
primary CMB and SZ cluster counts.  Moreover, the \emph{Planck} CMB
cosmology results are in tension with numerous independent large-scale
structure probes of cosmology in addition to cluster number counts
\citep[e.g.][]{Heymans13, Mandelbaum13, McCarthy14, Beutler14,
  Samushia14, Battaglia15a, Planck15SZps, Hojjati15}, adding further
indirect support to our results.  It has been suggested that the
\emph{Planck} CMB/clusters tension might point to exciting new
physics, including possible constraints on neutrinos
\citep[e.g.][]{Planck15SZcos}.  However, it is clear that significant
further work is first required on systematic uncertainties in cluster
mass measurement, especially for clusters at $z>0.3$.

\section*{Acknowledgments}

We acknowledge helpful discussions with Doug Applegate, Nick
Battaglia, Stefano Ettori, Henk Hoekstra, Anja von der Linden, Adam
Mantz, Jean-Baptiste Melin, and Mauro Sereno.  GPS acknowledges
support from the Royal Society.  GPS, FZ, CPH, SLM, ML, PEM
acknowledge support from the Science and Technology Facilities Council
(STFC).  This work was supported by ``World Premier International
Research Center Initiative (WPI Initiative)'' and the Funds for the
Development of Human Resources in Science and Technology under MEXT,
Japan, and Core Research for Energetic Universe in Hiroshima
University (the MEXT program for promoting the enhancement of research
universities, Japan).  CPH was funded by CONICYT Anillo project
ACT-1122.  MJ acknowledges the STFC [grant number ST/F001166/1].

\label{lastpage}

\end{document}